\documentclass[iop]{emulateapj}
\usepackage{graphicx}
\usepackage{amssymb}
\usepackage{epstopdf}
\usepackage{amsmath}
\usepackage{mathrsfs}
\usepackage{natbib}
\usepackage{setspace}
\usepackage{float}
\usepackage{times}
\defcitealias{Wen2013}{WH13}
\defcitealias{Cohen2014}{C14}

\slugcomment{Accepted for publication in The Astrophysical Journal}

\begin{document}

\shorttitle{Star Formation and Relaxation in Clusters}
\shortauthors{Cohen et al.}

\title{Star Formation and Relaxation in 379 Nearby Galaxy Clusters}

\author{Seth A. Cohen, Ryan C. Hickox, Gary A. Wegner}
\affil{Department of Physics and Astronomy, Dartmouth College, 6127 Wilder Laboratory, Hanover, NH 03755, USA}

\begin{abstract}
We investigate the relationship between star formation (SF) and level of relaxation in a sample of 379 galaxy clusters at $\textnormal{z} < 0.2$. We use data from the Sloan Digital Sky Survey to measure cluster membership and level of relaxation, and to select star-forming galaxies based on mid-infrared emission detected with the Wide-Field Infrared Survey Explorer.  For galaxies with absolute magnitudes $M_{r} < -19.5$, we find an inverse correlation between SF fraction and cluster relaxation: as a cluster becomes less relaxed, its SF fraction increases.  Furthermore, in general, the subtracted SF fraction in all unrelaxed clusters ($0.117 \pm 0.003$) is higher than that in all relaxed clusters ($0.097 \pm 0.005$).  We verify the validity of our SF calculation methods and membership criteria through analysis of previous work.  Our results agree with previous findings that a weak correlation exists between cluster SF and dynamical state, possibly because unrelaxed clusters are less evolved relative to relaxed clusters.
\end{abstract}


\section{Introduction}
\label{sec:IntroSec}

The study of galaxy cluster mergers on star formation (SF) has made significant progress in recent years. While the relationship between morphological type and clustercentric distance and local density has been well-known for decades \citep[e.g.,][]{Dressler1980}, mergers of clusters have been shown to affect these trends.  Many clusters exhibit an enhanced number of star-forming galaxies that authors attribute to the presence of substructure and thus to cluster merger activity \citep[e.g.,][]{Bird1993, Knebe2000}: for example, A98 and A115 \citep{Metevier2000}, A1367 \citep{Cortese2004}, A3921 \citep{Ferrari2005}, A3158 \citep{Johnston2008}, A85 \citep{Bravo2009}, RXCJ 0014.3-3022 \citep{Braglia2009}, A168 \citep{Hwang2009}, MACS J0025.4-1225 \citep{Ma2010}, and A2465 \citep{Wegner2011, Wegner2015}. In general, star formation rate (SFR) declines rapidly since $z \sim 2$, and several authors \citep[e.g.,][]{Popesso2012, Biviano2011, Koyama2010} have considered this for galaxy clusters as a function of redshift and cluster richness. \citet{Sobral2015} discuss significant boosting in SFR and AGN activity found in some merging clusters (e.g., CIZA J2242.8 +5301). Other studies report no effect from merging on cluster galaxies, and some suggest that SF is quenched by the interactions: for example, A168 \citep{Tomita1996}, A2356 \citep{Metevier2000}, post-starburst galaxies in A3921 \citep{Ferrari2005}, RXCJ 2308.3-0211 \citep{Braglia2009}, A1750 \citep{Hwang2009}, and A1664 \citep{Kleiner2014}. Modelling of mergers by \citet{Vijayaraghavan2013} indicate that quenching is important.

Recently, \citet[][hereafter C14]{Cohen2014} compared the SF and substructure properties of 107 clusters at $z < 0.1$ using optical spectroscopic data from the Sloan Digital Sky Survey (SDSS) and substructure information from \citet{Einasto2012}.  In general, they found a weak correlation between the amount of substructure and fraction of star-forming galaxies in their cluster sample.  In our paper, we perform a similar study on a larger sample of galaxy clusters utilizing different methods of SF and substructure detection.  We analyze 379 clusters at $z < 0.2$ using data from the SDSS and the Wide-Field Infrared Survey Explorer (WISE) to calculate SF information, and substructure determinations from \citet[][hereafter WH13]{Wen2013}.  In \S\ref{sec:DataSec}, we describe our cluster sample and discuss methods of SF calculation and substructure determination.  We present our results in \S\ref{sec:ResultsSec} and explain various verification tests against the results of \citetalias{Cohen2014} in \S\ref{sec:VerificationSec}.  We discuss our interpretations in \S\ref{sec:DiscussionSec}.

Throughout our analysis we assume a standard cosmology of $H_{0} = 70\:\textnormal{km}\:\textnormal{s}^{-1}\:\textnormal{Mpc}^{-1}$, $\Omega_{\textnormal{m}} = 0.27$, and $\Omega_{\Lambda} = 0.73$.

\section{Data and Sample Selection}
\label{sec:DataSec}

\subsection{Sample and Completeness}
\label{sec:SampleSec}

Our galaxy cluster sample is taken from \citetalias{Wen2013}, who measure the relaxation states of 2092 clusters in the redshift range $0.05 \le z \le 0.42$.  These clusters are from the catalog of \citet{Wen2012}, a collection of 132,684 clusters from the SDSS DR8 \citep{Aihara2011}.  Our optical data is from the SDSS DR9 \citep{Ahn2012}, from which we obtain photometric and, when available, spectroscopic redshifts; \emph{ugriz} magnitudes; and \emph{K}-corrections.  Mid-infrared data for determining SF properties is from the AllWISE catalog \citep{Wright2010, Mainzer2011}, which supplies magnitudes in four bands centered at 3.4, 4.6, 12, and 22 $\mu$m (hereafter, W1, W2, W3, and W4).

To determine cluster membership, we begin by following the procedure described in \citetalias{Wen2013}, to which we direct the reader for details.  In short, they select member galaxies based on photometric absolute magnitude, redshift, and projected distance from the cluster center.  However, we found that this method includes many background galaxies found above the clusters' red sequence in $^{0.1}(u-r)$ versus $M_{r}^{e}$ space, where $M_{r}^{e}$ is the evolution-correction $z = 0.1$ \emph{r}-band absolute magnitude and the superscript 0.1 denotes a \emph{K}-correction to a redshift of 0.1.  We therefore remove all galaxies at $^{0.1}(u-r) > 4$, about 6.7\% of the galaxy sample, which is 1.5--2 magnitudes above the red sequence as we observe for our clusters and as identified by, for example, \citet{Lisker2008} or \citet{Barazza2009}.  We choose this method for background galaxy removal rather than a more sophisticated technique to remain as close as possible to the membership in \citetalias{Wen2013}.  As part of their membership determinations, \citetalias{Wen2013} calculate $r_{200}$, the cluster radius at which the density is 200 times the critical density, via a relation to total \emph{r}-band luminosity within 1 Mpc as described in \citet{Wen2012}.

Since our sample is flux-limited, we must correct for the fact that galaxies at a given luminosity are increasingly difficult to detect at increasing redshift.  Thus, we include only galaxies brighter than a certain limit in $M_{r}^{e}$, which is determined by the absolute magnitudes of the faintest galaxies seen at the highest redshift of our sample.  However, above $z \approx 0.2$, imposing this absolute magnitude limit eliminates too many star-forming (and therefore generally fainter) galaxies for a SF analysis to be effective.  Therefore, we limit our sample of clusters to those at $z < 0.2$ and galaxies to those with $M_{r}^{e} < -19.5$. At $z = 0.2$, this limit corresponds to an apparent magnitude of approximately 20.4. We further restrict our sample to only those clusters with a W3 completeness level of at least 80\%, as explained in greater detail in \S\ref{sec:SFSec}. These cuts result in a final sample of 379 clusters. Within $r_{200}$, these clusters contain 40,792 galaxies, of which 7371 have spectroscopic redshifts; within $3r_{200}$, the clusters contain 69,980 galaxies, of which 17,726 have spectroscopic redshifts.  All galaxies in our sample have photometric redshifts.

\begin{figure}
\begin{center}
\includegraphics[scale=0.68, trim = 0.25in -0.02in 0.8in 0.3in, clip]{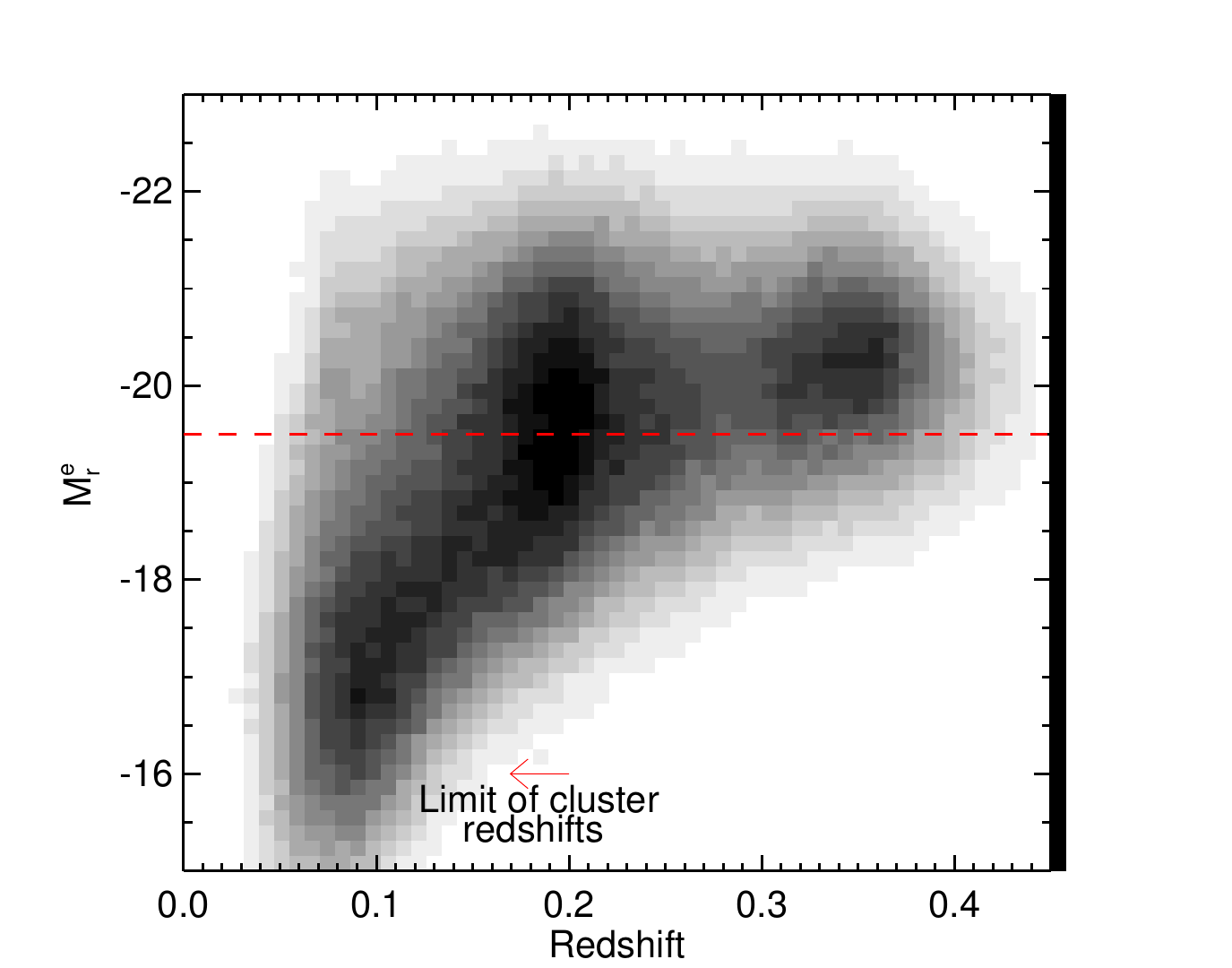}\llap{\raisebox{0.82cm}{\includegraphics[scale=0.25]{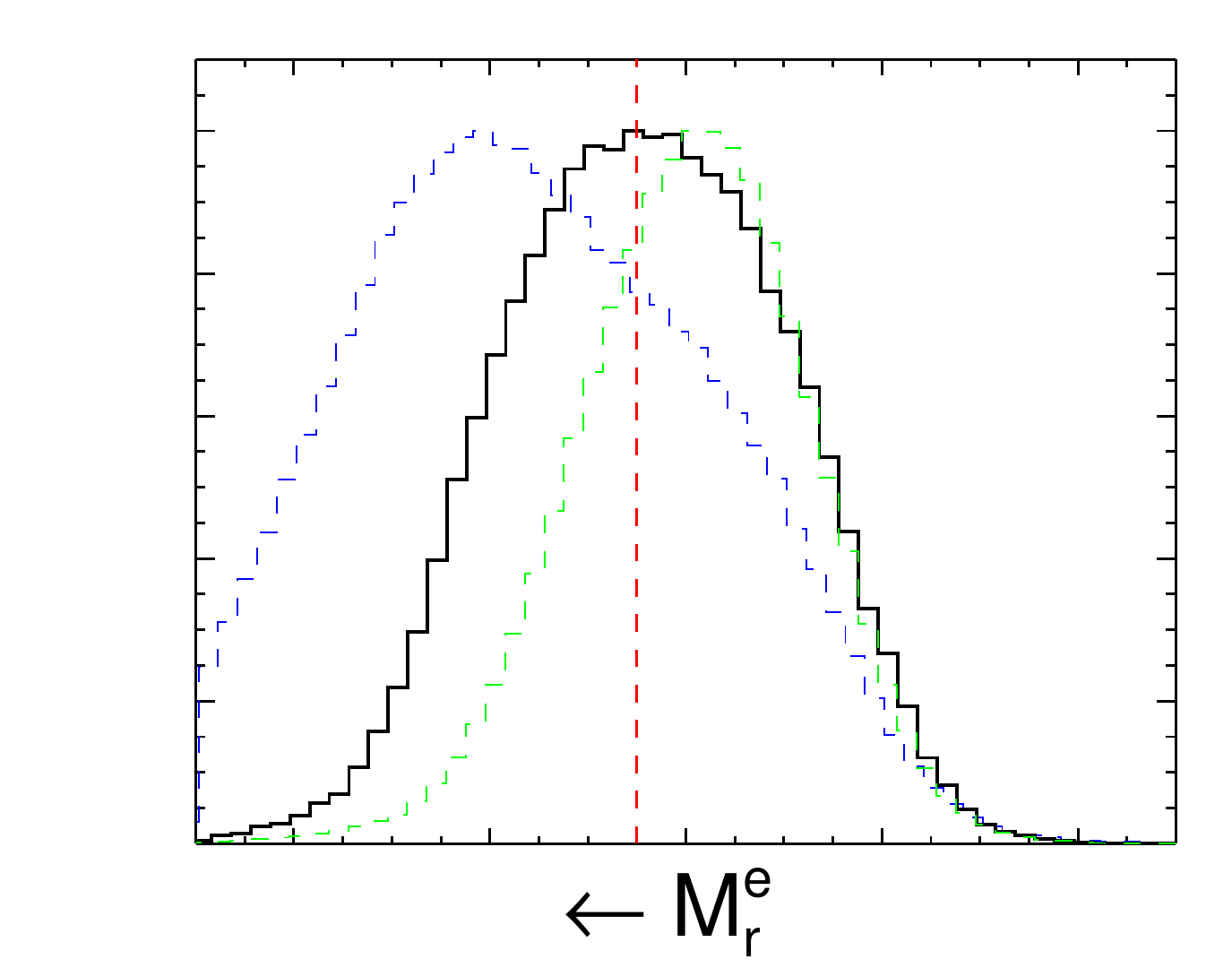}}}
\caption{$M_{r}^{e}$ versus redshift for galaxies with $M_{r}^{e} < -15$.  For clusters at our redshift limit of $z = 0.2$ (red arrow), our sample is complete to $M_{r}^{e} = -19.5$, as shown by the red dashed line.  \emph{Inset}: Histograms of $M_{r}^{e}$ for galaxies at $z < 0.18$ (blue), $0.18 < z < 0.2$ (black), and $z > 0.2$ (green).  The arrow in the x-axis label indicates the direction of increasing $M_{r}^{e}$.  Near $z = 0.2$, the vertical dashed line demarcates our completeness limit of $M_{r}^{e} = -19.5$.}
\label{fig:MagCutFig}
\end{center}
\end{figure}

Our galaxy sample is illustrated in Figure~\ref{fig:MagCutFig}, which shows $M_{r}^{e}$ versus redshift for galaxies with $M_{r}^{e} < -15$.  At the redshift limit of our clusters, $z = 0.2$, our sample of galaxies is complete to $M_{r}^{e} = -19.5$, as shown by the dashed line.  This is further illustrated in the inset, which shows histograms of $M_{r}^{e}$ for galaxies in various redshift bins.  Near $z = 0.2$, the completeness of our sample begins to drop at $M_{r}^{e} \approx -19.5$ (dashed line).  We note that galaxies at lower and higher redshifts are complete to fainter and brighter magnitudes, respectively.  Small adjustments to our completeness limit do not affect our conclusions.

\subsection{Measurements of Star-Forming Fraction}
\label{sec:SFSec}

We identify active galaxies via their detection in the WISE 12 $\mu$m band; we select objects with a signal-to-noise ratio greater than 3 as defined in the WISE data processing pipeline.  However, more nearby clusters contain a higher fraction of 12 $\mu$m-detected galaxies than distant clusters due to lower luminosity limits.  To correct for this, we restrict our definition of detection to include only those galaxies whose 12 $\mu$m luminosities ($\textnormal{L}_{12\mu\textnormal{m}}$) are complete across all redshifts of our sample.  We determine $\textnormal{L}_{12\mu\textnormal{m}}$ using the SED templates and codes of \citet{Chary2001}, modified to calculate the rest-frame flux in W3\footnote{We use a W3 zero-magnitude flux density of $F_{\nu^{0}} = 31.674 \:\textnormal{Jy}$, from the WISE data processing website, updated 2012 August 20.} \citep[e.g.,][]{Rosario2013, Webb2013}. This flux is further used below to calculate rest-frame W3 magnitudes.

\begin{figure}
\begin{center}
\includegraphics[scale=0.69, trim = 0.28in 0.04in 0.87in 0.3in, clip]{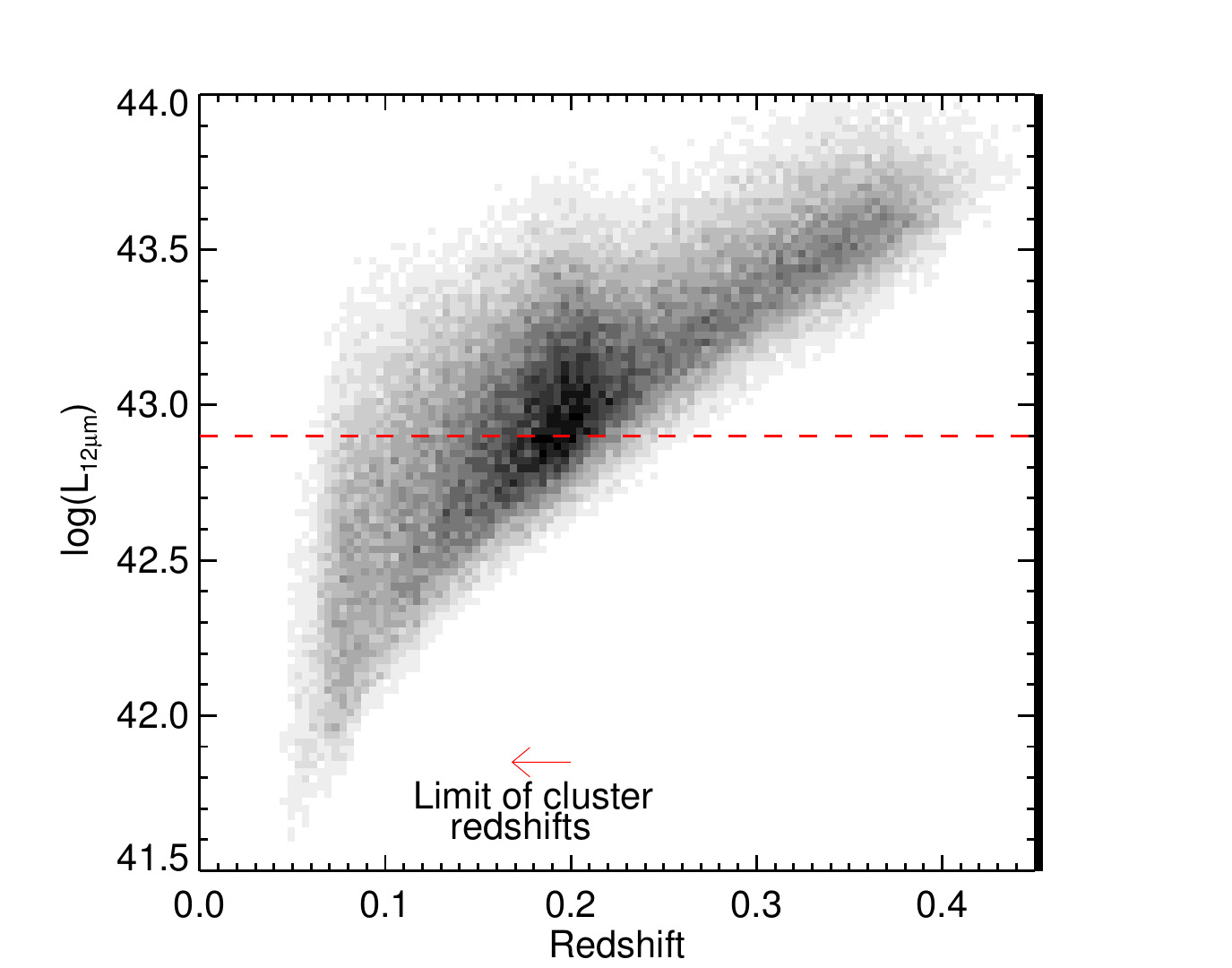}\llap{\raisebox{0.81cm}{\includegraphics[scale=0.25]{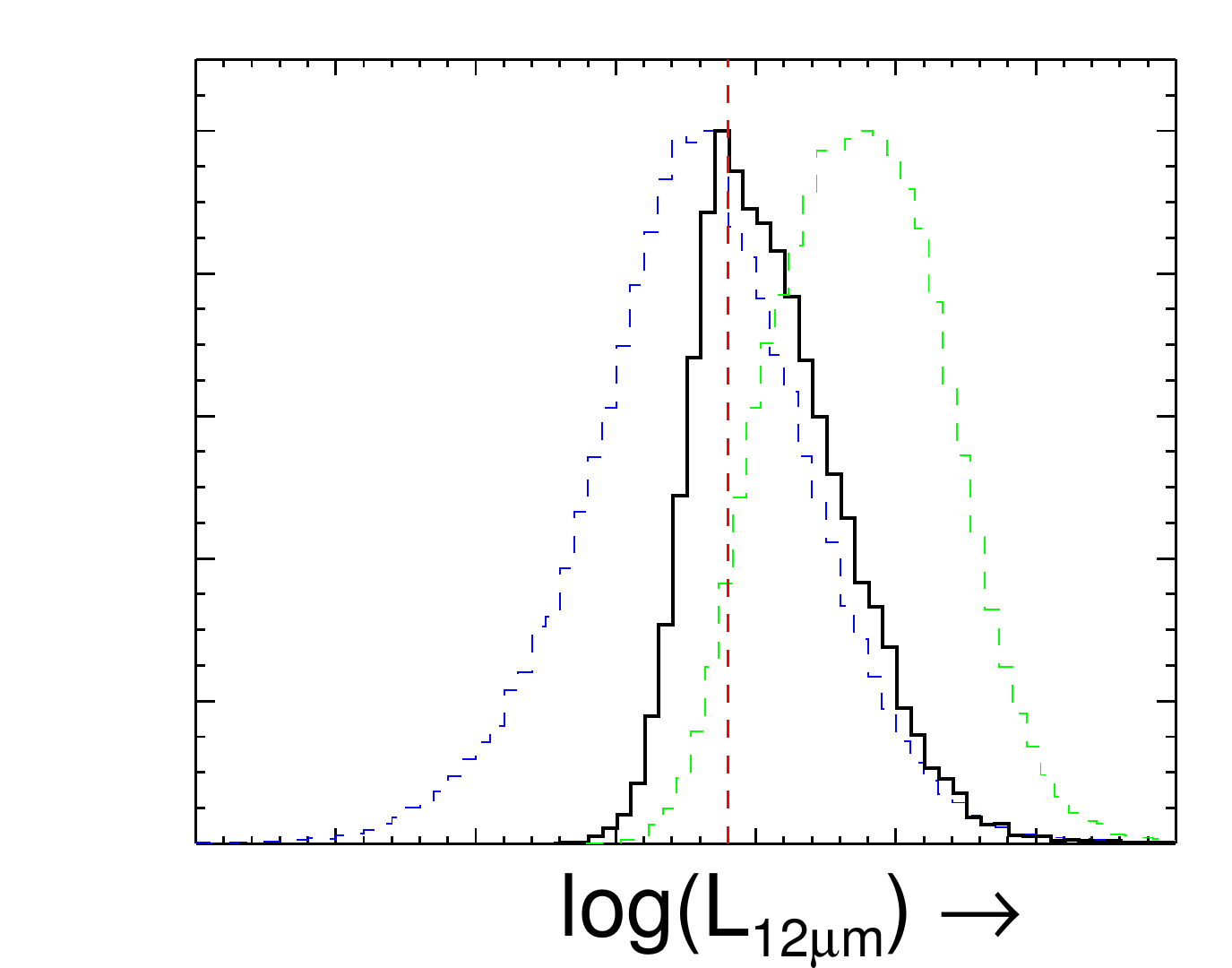}}}
\caption{$\log(\textnormal{L}_{12\mu\textnormal{m}})$ versus redshift for galaxies in our sample.  For clusters at our redshift limit of $z < 0.2$ (red arrow), we restrict our definition of detection in W3 to $\log(\textnormal{L}_{12\mu\textnormal{m}}) > 42.9$, as shown by the red dashed line.  \emph{Inset}: Histograms of $\log(\textnormal{L}_{12\mu\textnormal{m}})$ for galaxies at $z < 0.18$ (blue), $0.18 < z < 0.2$ (black), and $z > 0.2$ (green).  The arrow in the x-axis label indicates the direction of increasing $\log(\textnormal{L}_{12\mu\textnormal{m}})$.  Near $z = 0.2$, the vertical dashed line demarcates our completeness limit of $\log(\textnormal{L}_{12\mu\textnormal{m}}) = 42.9$.}
\label{fig:LumCutFig}
\end{center}
\end{figure}

As illustrated in Figure~\ref{fig:LumCutFig}, at the redshift limit of our clusters, $z = 0.2$, our sample of galaxies is complete to $\log(\textnormal{L}_{12\mu\textnormal{m}}) \approx 42.9$, which we use as our $\textnormal{L}_{12\mu\textnormal{m}}$ limit.  As with Figure~\ref{fig:MagCutFig}, the histograms of $\log(\textnormal{L}_{12\mu\textnormal{m}})$ in the inset further illustrate this completeness limit, which corresponds to a star formation rate (SFR) of approximately $2.3\:M_{\odot}\:yr^{-1}$ (using the relation in \citealt{Donoso2012}) or approximately $3.7\:M_{\odot}\:yr^{-1}$ (using the relation in \citealt{Lee2013}).  Using slightly different $\textnormal{L}_{12\mu\textnormal{m}}$ limits does not affect our conclusions.

This $\textnormal{L}_{12\mu\textnormal{m}}$ limit is further complicated by the internal WISE completeness as a function of flux density. At $z = 0.2$, our limit of $\log(\textnormal{L}_{12\mu\textnormal{m}}) = 42.9$ corresponds to a W3 flux density of approximately $670\:\mu\textnormal{Jy}$.  For the $> 90\%$ of our galaxies whose WISE depth is at least 11 frames, this W3 flux density corresponds to a completeness of at least 90\%. Additionally, over half of these galaxies have a depth greater than 14 frames, corresponding to a completeness of greater than 95\%. Despite this high completeness, we still correct the number of star-forming galaxies in our sample to account for the missing galaxies. To do this, we adjust the number of detected star-forming galaxies by the percent completeness at each galaxy's flux density, determined via interpolation of the internal W3 completeness curves.\footnote{WISE completeness curves are found on the WISE data processing website, updated 2012 March 16.} To avoid relying too heavily on this completeness correction, we remove from our sample all clusters containing any galaxies whose flux density corresponds to a W3 completeness of less than 20\%.

To distinguish between star-forming galaxies and AGN, many past studies have employed cuts in WISE color space \citep[e.g.,][]{Jarrett2011, Stern2012, Assef2012, Yan2013, Satyapal2014}.  \citet{Stern2012}, for example, defined AGN as those galaxies with $\textnormal{W}1 - \textnormal{W}2 > 0.8$, while \citet{Satyapal2014} investigated a more liberal cut of $\textnormal{W}1 - \textnormal{W}2 > 0.5$.  In this work, we define star-forming galaxies as those with rest-frame WISE colors $\textnormal{W}1 - \textnormal{W}2 < 0.6$.  Adjusting this threshold does not affect our conclusions.

\begin{figure}
\begin{center}
\includegraphics[scale=0.63]{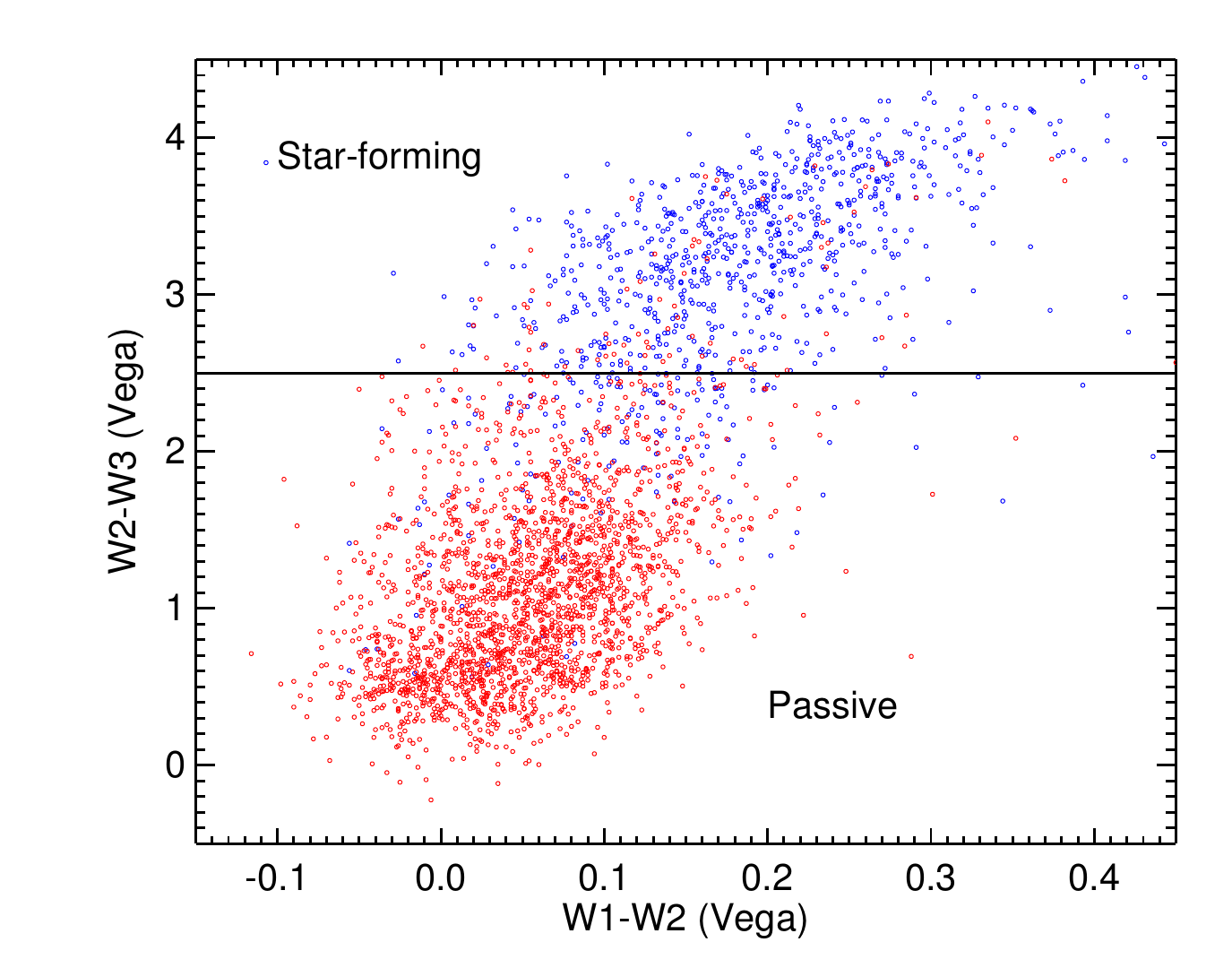}
\caption{$\textnormal{W}2 - \textnormal{W}3$ versus $\textnormal{W}1 - \textnormal{W}2$ for galaxies from \citetalias{Cohen2014}.  Blue and red points indicate star-forming galaxies and AGN, respectively, as determined via optical spectroscopy.  The horizontal line at $\textnormal{W}2 - \textnormal{W}3 = 2.5$ marks our $\textnormal{W}2 - \textnormal{W}3$ color cut.}
\label{fig:W23vsW12Fig}
\end{center}
\end{figure}

To separate star-forming and passive galaxies, we define star-forming galaxies as those with rest-frame $\textnormal{W}2 - \textnormal{W}3 > 2.5$.  This limit is chosen using the cluster sample of \citetalias{Cohen2014}, as illustrated in Figure~\ref{fig:W23vsW12Fig}, which shows $\textnormal{W}2 - \textnormal{W}3$ versus $\textnormal{W}1 - \textnormal{W}2$ for the galaxies from \citetalias{Cohen2014}.  Blue and red points indicate star-forming and passive galaxies, respectively, as determined in \citetalias{Cohen2014} via optical spectroscopy.  At $\textnormal{W}2 - \textnormal{W}3 > 2.5$, almost 90\% of the galaxies are star-forming.

We calculate rest-frame W1 and W2 magnitudes by assuming that these bands are dominated by emission from the Rayleigh-Jeans tail of the galaxies' SEDs.  Since Rayleigh-Jeans flux is proportional to $\nu^{2}$, we correct the measured W1 and W2 fluxes by $1/(1+z)^{2}$ to determine rest-frame fluxes and magnitudes.

When calculating the SF fraction for each cluster, we take into account the fact that the cluster membership procedure of \citetalias{Wen2013} imposes liberal limits on their photometric redshifts (photo-\emph{z}s) that include a number of foreground and background galaxies that are not cluster members.  Indeed, for each cluster, \citetalias{Wen2013} include galaxies within a photo-\emph{z} slice of $z \pm 0.04(1+z)$.  For a cluster at $z = 0.1$, for example, this corresponds to a comoving distance of over 300 Mpc between the close and far edges of the cluster.

To correct for the large width of this redshift bin, we estimate the number of foreground and background galaxies included as cluster members, and subtract these galaxies as part of our SF fraction calculation.  To do this, we assume that the field just outside, and in the same redshift range as, the clusters will contain comparable numbers of foreground and background galaxies as the clusters' lines of sight.  We calculate SF fraction within both $r_{200}$ and $3r_{200}$, and the region outside a cluster is defined as between $3r_{200}$ and $5r_{200}$.  We define subtracted SF fraction as

\begin{equation}
\label{equ:NormSFfracEqu}
\textnormal{Subtracted SF Fraction} = \frac{N^{in}_{SF} - N^{out}_{SF}A_{ratio}}{N^{in}_{all} - N^{out}_{all}A_{ratio}},
\end{equation}
where $N$ is number of galaxies; the superscripts denote galaxies inside ($in$) or outside ($out$) the cluster; the subscripts distinguish between number of star-forming galaxies ($SF$) or total number of galaxies ($all$); and the multiplication by $A_{ratio}$ normalizes the area outside the cluster to the area inside.  $N_{SF}$ is corrected for internal W3 completeness, while $N_{all}$ is not. The galaxies outside the clusters are selected in the same way as the cluster galaxies.  This subtraction, then, effectively cancels the potential contamination of the SF fraction caused by foreground and background galaxies.

\subsection{Relaxation Measurements}
\label{sec:RelaxSec}

To quantify their clusters' dynamical states, \citetalias{Wen2013} assign each cluster a relaxation parameter $\Gamma$ via three symmetry tests of the smoothed \emph{r}-band surface brightness maps of the galaxies within $r_{500} = 2/3\;r_{200}$.  We explain these tests briefly here; for more details, see \citetalias{Wen2013}.  First, the asymmetry factor $\alpha$ quantifies the rotational symmetry of a cluster.  Second, the ridge flatness $\beta$ utilizes the radial light profile steepness in many angular directions, with unrelaxed clusters exhibiting flatter profiles.  Finally, the normalized deviation $\delta$ quantifies the smoothed optical map's deviation from the two-dimensional elliptical King model.

To determine the final relaxation parameter $\Gamma$ of each cluster, \citetalias{Wen2013} first define a plane in the three-dimensional space of $\alpha$, $\beta$, and $\delta$ that optimizes the separation between relaxed and unrelaxed clusters as determined by X-ray imaging.  $\Gamma$ is defined as the distance from this plane.  Positive values of $\Gamma$ indicate relaxed clusters, while negative values of $\Gamma$ denote unrelaxed clusters.  \citetalias{Wen2013} show that this relaxation parameter is reasonably well correlated with dynamical parameters of clusters derived from X-ray data, such as concentration, centroid shift, power ratio, and cooling time.

\section{Results}
\label{sec:ResultsSec}

\begin{figure}
\begin{center}
\includegraphics[scale=0.63]{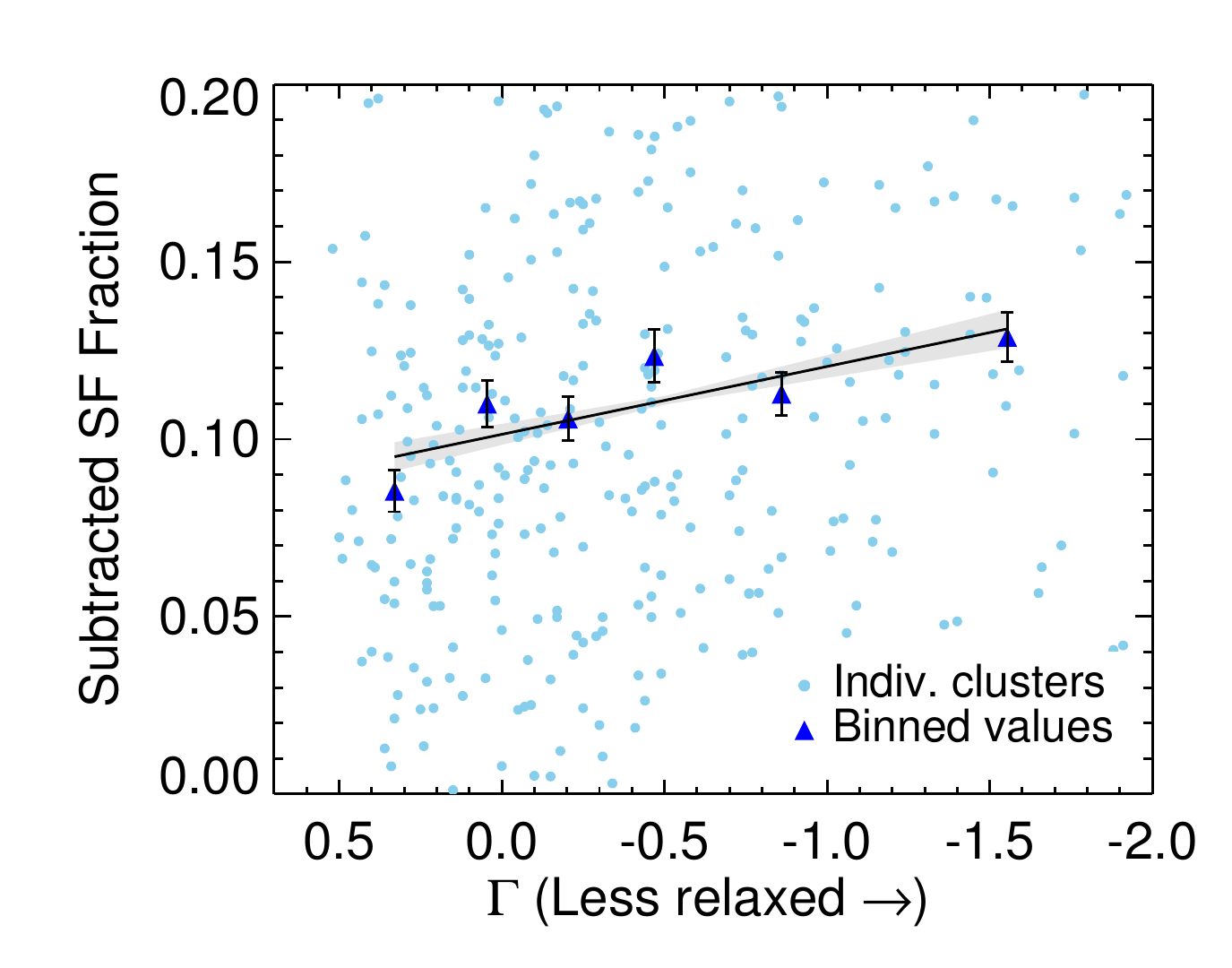}
\caption{Subtracted SF fraction versus $\Gamma$, the relaxation parameter from \citetalias{Wen2013}.  Relaxation decreases to the right.  Light blue points represent individual clusters, and dark blue triangles are the total SF fractions of all galaxies in each bin.  The gray region represents a $1\sigma$ error on the best fit solid line.  For clarity, we have not plotted $<10\%$ of clusters whose SF fractions are slightly above 0.2 and whose $\Gamma$ values span the plotted range, and the legend box obscures a small number of points.  In general, less relaxed clusters exhibit more SF.}
\label{fig:SFfracvsGammaFig}
\end{center}
\end{figure}

In Figure~\ref{fig:SFfracvsGammaFig}, we plot the subtracted SF fraction within $3r_{200}$, as discussed in \S\ref{sec:SFSec}, as a function of relaxation parameter $\Gamma$.  Notice that $\Gamma$, and thus relaxation, decreases to the right.  Each light blue point represents a cluster.  The dark blue triangles represent the SF fraction of all cluster galaxies in each bin in $\Gamma$, and the errors on these points are calculated via a bootstrap resampling of the galaxies in each bin.  Each bin measures the total SF fraction of all galaxies in that bin.  The grey region represents a $1\sigma$ error on the best fit solid line to the binned values, which is calculated by minimizing the chi-square error statistic of the data.

The slope of the relation is $0.020 \pm 0.004$, a significance of about $4.4\sigma$, which is calculated as the formal uncertainty on the linear chi-square fit.  This indicates that a weak but significant correlation exists between SF in and decreased relaxation of clusters.  This correlation is confirmed by Spearman's rank test: the binned values produce a correlation with $\rho = 0.89$ and $P = 0.019$, indicating a strong, significant correlation.  Additionally, Spearman's test on the individual cluster values produces a correlation with $\rho = 0.17$ and $P = 0.001$, indicating a weak but significant correlation.

As another test of this relationship between SF and relaxation, we also calculate the total SF fraction in all relaxed clusters (those with $\Gamma > 0$) and in all unrelaxed clusters (those with $\Gamma < 0$).  We find that the subtracted SF fraction in unrelaxed clusters, $0.117 \pm 0.003$, is higher than the SF fraction in relaxed clusters, $0.097 \pm 0.005$.  The significance of this difference between unrelaxed and relaxed clusters is approximately $3.6\sigma$.  For comparison, the SF fraction for all field galaxies at $3r_{200} < r < 5r_{200}$ is approximately $0.221 \pm 0.003$. Over this large sample of clusters, then, a more unrelaxed state is correlated with higher cluster SF.

As discussed in \S\ref{sec:SFSec}, we also perform this analysis only using galaxies within $r_{200}$.  We find a slope of $0.018 \pm 0.003$ and subtracted SF fractions of $0.114 \pm 0.002$ for unrelaxed clusters and $0.095 \pm 0.003$ for relaxed clusters.  These results are similar in value and significance to those above and produce the same conclusions.  Furthermore, the observed correlation remains when using the means or medians of the cluster SF fractions in each bin. The significance of the correlation is slightly lower, but this could be due to other factors that affect SF in clusters (e.g., merger history, as discussed in \S\ref{sec:DiscussionSec}.)

\section{Verification}
\label{sec:VerificationSec}

In this section, we discuss several tests we perform to check the validity of our SF calculation methods, and to compare our results to those in \citetalias{Cohen2014}, who found that the SF fraction in clusters with substructure, $0.228 \pm 0.007$, is higher than that in clusters without substructure, $0.175 \pm 0.016$.  This result agrees qualitatively with ours, and we discuss the implications of this in \S\ref{sec:DiscussionSec}.  However, the SF fractions in our unrelaxed and relaxed clusters are lower, and the absolute difference between these fractions is smaller, than those in \citetalias{Cohen2014}.  In the following, we argue that these differences are expected and a consequence of the membership selection and SF identification methods used in the current paper.

All results from this paper, \citetalias{Cohen2014}, and the verification tests discussed below are summarized in Table~\ref{tab:DataTab}, with the following columns: (1) source of cluster sample; (2) database and method from which SF information is calculated; (3) number of galaxies; (4), (5), \& (6) redshift, absolute magnitude, and $\log(\textnormal{L}_{12\mu\textnormal{m}})$ limits, respectively; (7) whether cluster membership is determined spectroscopically or photometrically; (8) radius within which SF fraction is determined; (9) \& (10) SF fractions for unrelaxed (multi-component) and relaxed (one-component) clusters, respectively; and (11) SF fractions for field galaxies in the cluster region, defined as being at $3r_{200} < r < 5r_{200}$.  Note that in all verification tests, we correct the number of detected star-forming galaxies using WISE completeness curves, and we report subtracted SF fractions, as explained in \S\ref{sec:SFSec}.  For reference, the first two rows display the main results from the current work, and the last row displays the results from \citetalias{Cohen2014}.

\begin{deluxetable*}{cccccccccccc}
\tablecolumns{12}
\tablewidth{0pc}
\tabletypesize{\scriptsize}
\tablehead{
\colhead{} & \colhead{} & \colhead{} & \colhead{} & \multicolumn{3}{c}{Completeness Limits} & \colhead{} & \colhead{} & \multicolumn{3}{c}{SF Fractions} \\
\cline{5-7} \cline{10-12} \\
\colhead{} & \colhead{\textbf{Sample}} & \colhead{\textbf{SF Method}} & \colhead{$N_{gals}$} & \colhead{\textbf{z}} & \colhead{\textbf{$M^{e}_{r}$}} & \colhead{$\log(\textnormal{L}_{12\mu\textnormal{m}})$} & \colhead{\textbf{Memb.}} & \colhead{\textbf{Radius}} & \colhead{\textbf{Unrelaxed}\tablenotemark{a}} & \colhead{\textbf{Relaxed}\tablenotemark{b}} & \colhead{\textbf{Field}\tablenotemark{c}} \\
\colhead{} & \colhead{(1)} & \colhead{(2)} & \colhead{(3)} & \colhead{(4)} & \colhead{(5)} & \colhead{(6)} & \colhead{(7)} & \colhead{(8)} & \colhead{(9)} & \colhead{(10)} & \colhead{(11)}}
\startdata
(1) & This work & WISE & 69980 & 0.2 & -19.5 & 42.9 & Phot\tablenotemark{d} & $3r_{200}$ & $0.117 \pm 0.003$ & $0.097 \pm 0.005$ & $0.221 \pm 0.003$ \\
(2) & This work & WISE & 40792 & 0.2 & -19.5 & 42.9 & Phot\tablenotemark{d} & $r_{200}$ & $0.114 \pm 0.002$ & $0.095 \pm 0.003$ & --- \\
\hline \hline
(3) & \citetalias{Cohen2014} & WISE & 7456 & 0.1 & -20.5 & 42.9 & Phot & $3r_{200}$ & $0.086 \pm 0.006$ & $0.061 \pm 0.013$ & $0.144 \pm 0.006$ \\
(4) & \citetalias{Cohen2014} & WISE & 6037 & 0.1 & -20.5 & 42.9 & Spec & $3r_{200}$ & $0.096 \pm 0.007$ & $0.068 \pm 0.014$ & $0.166 \pm 0.007$ \\
\hline
(5) & \citetalias{Cohen2014} & WISE & 2717 & 0.1 & -20.5 & 42.9 & Phot & $r_{200}$ & $0.070 \pm 0.007$ & $0.065 \pm 0.013$ & --- \\
(6) & \citetalias{Cohen2014} & WISE & 2224 & 0.1 & -20.5 & 42.9 & Spec & $r_{200}$ & $0.079 \pm 0.007$ & $0.071 \pm 0.014$ & --- \\
\hline \hline
(7) & This work & WISE & 8359 & 0.1 & -20.5 & 42.1 & Phot & $3r_{200}$ & $0.193 \pm 0.012$ & $0.185 \pm 0.015$ & $0.404 \pm 0.009$ \\
(8) & \citetalias{Cohen2014} & SDSS & 4151 & 0.1 & -20.5 & --- & Spec & $3r_{200}$ & $0.228 \pm 0.007$ & $0.175 \pm 0.016$ & ---
\enddata
\tablenotetext{a}{In \citetalias{Cohen2014}, ``Unrelaxed" refers to multi-component clusters.}
\tablenotetext{b}{In \citetalias{Cohen2014}, ``Relaxed" refers to one-component clusters.}
\tablenotetext{c}{Field galaxies are defined as being at $3r_{200} < r < 5r_{200}$.}
\tablenotetext{d}{As explained in \S\ref{sec:SampleSec}, while we use both photometric and spectroscopic data to identify cluster members, most galaxies are photometrically selected.}
\tablecomments{Results from this work (rows 1 and 2), \citetalias{Cohen2014} (row 8), and several verification tests.  For fair comparison, all tests are performed using the absolute magnitude and redshift limits imposed in \citetalias{Cohen2014}, $M_{r}^{e} < -20.5$ and $z < 0.1$.  Agreement among this work's results and the various verification tests demonstrates the validity of our SF calculation methods.}
\label{tab:DataTab}
\end{deluxetable*}

In our two main verification tests, we re-calculate the SF fractions of clusters with and without substructure from the cluster sample of \citetalias{Cohen2014}, but use WISE data to classify a galaxy as star-forming, as in the current paper.  In one test, we select member galaxies using only photometric data as in \citetalias{Wen2013} (rows 3 and 5); in the other, we include only those galaxies detected spectroscopically (rows 4 and 6). In both cases, we examine galaxies within both $3r_{200}$ and $r_{200}$.  These tests allow us to directly compare different methods of SF detection and membership selection using the same cluster sample.  To ensure fair comparison, all tests are calculated with the absolute magnitude and redshift limits used in \citetalias{Cohen2014}, $M_{r}^{e} < -20.5$ and $z < 0.1$.

Several comparisons of these results are instructive.  First, we focus on the tests in rows 3 through 6.  At both cluster radii, we find statistically similar SF fractions between tests employing both spectroscopic and photometric membership methods, confirming that these methods achieve similar results.  We also note that the SF fractions within $r_{200}$ are lower than those within $3r_{200}$, since the central regions of clusters contain fewer star-forming galaxies (e.g., \citealt{Rines2005}; \citetalias{Cohen2014}).

Next, we note that the results from these verification tests are less significant than the main results of this work due to the smaller number of galaxies in the \citetalias{Cohen2014} sample and to the stricter absolute magnitude and redshift limits.  Additionally, comparing the results using the two membership selection methods shows how applying photometric membership criteria results in lower significance than using spectroscopic membership methods.  This illustrates the advantage of gathering a large cluster sample when utilizing photometric data, as we have done in this work.

The SF fractions from this work and from the discussed verification tests are lower than those from \citetalias{Cohen2014}, summarized in row 8.  This is due to the use of both WISE data for SF calculations and the relatively high $\textnormal{L}_{12\mu\textnormal{m}}$ limit necessary for a fair discussion of the comparisons above.  To illustrate this point, we perform the verification test summarized in row 7, which uses a lower limit of $\log(\textnormal{L}_{12\mu\textnormal{m}}) > 42.1$.  This limit is determined from the current work's galaxies at $z < 0.1$ and is complete to this redshift.  As expected, the resulting SF fractions are much more similar to those from \citetalias{Cohen2014}.

Finally, we note that the field SF fractions (column 11) differ for different cluster samples for two reasons. One, optically-brighter galaxies, like those included in the tests in rows 3 through 6, are less likely to be star-forming. Two, the lower luminosity limit used in the test in row 7 selects more star-forming galaxies.

These verification tests confirm that the methods used in this work -- utilizing WISE data in a photometrically-selected sample of galaxies to calculate subtracted SF fractions -- produce conclusions consistent with the methods used in \citetalias{Cohen2014}, which are based on more robust spectroscopic determinations of SF and cluster membership, but for a much smaller sample of clusters.

\section{Discussion}
\label{sec:DiscussionSec}

We find a higher fraction of star-forming galaxies in less-relaxed clusters than more-relaxed clusters.  This result agrees with the findings of \citetalias{Cohen2014}, who also found a correlation between cluster SF and cluster dynamical state.  This is especially promising because these studies measure SF and cluster relaxation with independent methods.  In particular, to determine SF properties, \citetalias{Cohen2014} used optical spectroscopic data from SDSS, while our study uses infrared data from WISE.  Furthermore, \citetalias{Cohen2014} measured substructure out to several virial radii using two- and three-dimensional statistical tests from \citet{Einasto2012}, while our study uses surface brightness symmetry tests out to $r_{500}$ from \citetalias{Wen2013}.  The fact that both studies arrive at the same conclusion strengthens the result that, in general, more dynamically active clusters exhibit higher amounts of SF.

As in \citetalias{Cohen2014}, we propose two possible explanations for these results.  First, unrelaxed clusters could exhibit higher SF fractions because the cluster dynamics causing the clusters to appear unrelaxed (i.e., merging) could be actively enhancing cluster SF.  However, we prefer a second explanation, that unrelaxed clusters are still in the process of forming and thus represent a transitional state between the field environment and a relaxed cluster environment.  Since field galaxies, in general, exhibit higher SF than cluster galaxies, a transitional state could exhibit SF values between those of these two environments.

A possible avenue for distinguishing between these explanations and for decreasing the large scatter in our observed SF fractions involves determining merger histories of our clusters, since clusters at different stages of merging can exhibit different SF fractions for similar apparent relaxation states \citep[e.g.,][]{Hwang2009}.  Analytical calculations (e.g., the radial infall model of \citealt{Beers1982}) and simulations utilizing clusters' velocities and masses \citep[e.g.,][]{Dawson2013, Poole2008} can be used to estimate merger histories of many clusters. Since these methods require knowledge of the masses of the substructures in each cluster, studies such as \citet{Parekh2015}, \citet{Einasto2012}, or \citet{Andrade2012} could provide useful cluster samples for this analysis.

\acknowledgements

We thank the referee for helpful suggestions, Wen Zhonglue for very helpful explanations regarding his paper, and the SDSS and WISE teams for the publicly available data releases.  Funding for SDSS-III is provided by the Alfred P. Sloan Foundation, the Participating Institutions, the NSF, and the U.S. D.O.E. Office of Science. The SDSS-III web site is http://www.sdss3.org/.  This publication makes use of data products from WISE, a joint project of UCLA and JPL/Caltech, and NEOWISE, a project of JPL/Caltech. WISE and NEOWISE are funded by NASA.

\end{document}